\documentstyle[11pt]{article}

\catcode`@=11

\@addtoreset{equation}{section}
\catcode`@=12

\title{\bf Quantum Group Schr\"{o}dinger Field Theory}
\author{\\ \\ \\ Marcelo R. Ubriaco\\
 {\em Laboratory of Theoretical Physics}\\
{\em Department of Physics}\\
{\em  P. O. Box 23343}\\
{\em University of Puerto Rico, Rio Piedras}\\
{\em PR 00931-3343,  USA}}
\date{LTP-035-UPR\\April 1993}

\begin{document}
\vspace{0.5in}

\maketitle

\vspace{.3in}
\begin{abstract}
We show that a quantum deformation of quantum mechanics
given in a previous work is equivalent to quantum mechanics on
a nonlinear lattice with step size $\Delta x=~(1-q)x$.  Then,
based on this, we develop the basic
formalism of  quantum group Schr\"{o}dinger field theory
in one spatial quantum  dimension, and explicitly exhibit the $SU_{q}(2)$
covariant algebras satisfied by the $q$-bosonic and $q$-fermionic
Schr\"{o}dinger fields.  We generalize this result to an arbitrary number of
fields.
\end{abstract}
\newpage

\section{Introduction}
The remarkable relation between quantum groups
and non-commutative geometry \cite{Woronowicz}\cite{Manin}, and its realization
in terms of non-commutative differential calculus \cite{WessZumino}
has recently  acquired a noticeable place in the physics literature.
The fact that non-commutative calculus can be understood as a
deformation of ordinary calculus served as a paradigm to study
$q$-deformations of standard physical systems.
In particular,  different aspects of quantum deformations
of quantum mechanics have been studied
by several authors \cite{Minahan}\cite{Wess}\cite{Ubriaco}.
In this paper, based on work in \cite{Ubriaco}, we study Schr\"{o}dinger
quantum group field theory in one spatial quantum dimension.
The main objective of this work is to develop
a formulation of a quantum deformed second quantized field theory,
which could set a basis to an analytical approach to a
field theory with higher dimensional quantum space-time symmetries.
In Section \ref{QM} we
briefly discuss quantum mechanics for $0<q<1$ and show the
equivalence between the corresponding deformed Schr\"{o}dinger hamiltonian
with a difference equation on a nonlinear lattice. In Section \ref{QFD}
we develop the basic formalism of a Schr\"{o}dinger
quantum field theory satisfying quantum commutation relations on
the nonlinear lattice, and
in Section \ref{fa} we obtain explicitly the $SU_{q}(2)$
covariant field algebras satisfied by $q$-bosonic and
$q$-fermionic Schr\"{o}dinger fields. We generalize this result
to an arbitrary number of fields and conclude with some remarks.
\section{Quantum Mechanics with $0<q<1$} \label{QM}
Given $q$ real, $0<q<1$, the  following deformation of the quantum Heisenberg
algebra
\begin{equation}
px-qxp=-i ,  \label{ps}
\end{equation}
is seen to be compatible with non-commutative differential calculus
on the quantum line once one defines $p\equiv -i\hat{\partial}$,
where $\hat{\partial}$ is the quantum derivative.
If we impose the hermiticity
condition on both $p$ and $x$,  then Equation (\ref{ps}) will
reduce to $[p,x]=~\frac{-2i}{q+1}$ and therefore
standard quantum mechanics.  Therefore, the only two alternatives
available is to require that either $p$ \cite{Wess} or $x$ \cite{Ubriaco}
be hermitian.
As discussed in \cite{Ubriaco}, by enforcing hermiticity on the
coordinate operator, $x= \overline{x}$, we can still keep the usual
meaning of  position,
speed, acceleration as real expectation values, allowing
besides to implement the formalism with $q$-differentials
and $q$-integrals.
It was found that one can
define a hermitian free hamiltonian $H_{0}$ as
\begin{equation}
H_{0}=g(q)p\overline{p}  \label{Ho}
\end{equation}
with $\overline{p}=q^{-1}T^{-1}(x)p$, $T(x)=q^{x\partial_{x}}$
and $g(q=1)=1$, such that
the time dependent deformed
Schr\"{o}dinger equation consistent with a time
independent probability density is given by
\begin{equation}
\left(-x^{-1}[x\partial_{x}]x^{-1}[x\partial_{x}]+V(q,x)\right)\Phi(x,t)
=i\partial_{t}\Phi(x,t)   \label{H}
\end{equation}
where $[x\partial_{x}]=\frac{T^{1/2}(x)-T^{-1/2}(x)}
{q^{1/2}-q^{-1/2}}$, and we have chosen $g(q)=q^{\frac{3}{2}}$.
By introducing an independent variable $z$ and considering
the coordinate $x$ as a function
$x(z)=q^{z}$,  Equation (\ref{Ho}) can be rewritten as a equation
taking values on a nonlinear lattice of variable step size
\begin{equation}
\Delta x(z)\equiv x(z)-x(z+1)=(1-q)x
\end{equation}
 Similarly, the difference operation on any function  $y(x(z))$ is
written
\begin{equation}
\Delta y(x(z))\equiv y(x(z))-y(x(z+1))=
\left(1-T(x)\right)y(x(z))
\end{equation}
such that on the lattice the free hamiltonian $H_{0}$ reads
\begin{equation}
H_{0}=-\frac{\Delta}{\nabla x_{1}}\left[\frac{\nabla y(x(z))}
{\nabla x(z)}\right]
\end{equation}
where $x_{1}=x(z+1/2)$ and $\nabla x(z)=\Delta x(z-1)$. Difference
equations with operators of this type were studied in \cite{Suslov} in the
context of the theory of basic hypergeometric functions.
\section{Quantum Field Deformations} \label{QFD}
In this section we look for an action that leads to Equation
(\ref{H}).  An obvious procedure is to take the usual Schr\"{o}dinger
action, replace the differential and integral
operators by their $q$-analogs and postulate the action
\begin{equation}
S=\int dt\int d_{q}x \left[- f(q) D_{x}^{(+)} \Phi^{\ast} D_{x}^{(+)}\Phi
+i  \Phi^{\ast} \dot{\Phi} - V(x)\Phi^{\ast} \Phi\right]
\end{equation}
where the $q$-derivative $D_{x}^{(+)}\equiv x^{-1}\frac{1-T(x)}{1-q}$ and the
$q$-integral operator is given by the Jackson integral
\cite{Jackson}
\begin{equation}
\int d_{q}x\:f(x)\equiv (1-q)x \sum_{n=0}^{\infty} q^{n}
f\left(q^{n}x\right)
\end{equation}
{}From the definition of
the inner product introduced in \cite{Ubriaco}
\begin{equation}
<\Psi,\Phi>\equiv\int d_{q}x \,{\Psi}^{\ast}\Phi=
(1-q)x\sum_{n=0}^{\infty} q^{n} {\Psi}^{\ast}(q^{n}x)\Phi(q^{n}x),
\end{equation}
we recall that the hermitian adjoint of the $q$-derivative is given by
\begin{equation}
\overline{D}_{x}^{(+)}=-q^{-1}D_{x}^{(-)}.
\end{equation}
Integration by parts
with use of the $q$-analog of the Leibniz rule
\begin{equation}
D_{x}^{(+)}\left[\Phi\phi \right]=\left[D_{x}^{(+)}\Phi \right]T(x)\phi+\Phi
\left[D_{x}^{(+)}\phi \right]
\end{equation}
shows that the action $S$ is hermitian.
Variation of $S$ with respect to $\Phi^{\ast}$
leads to
\begin{eqnarray}
\delta S & = &\int dt\int d_{q}x \,\delta\Phi^{\ast}
\left[f(q)D_{x}^{(+)}T^{-1}(x)
D_{x}^{(+)}\Phi+i\dot{\Phi}-V\Phi \right] \nonumber\\
 &  & -\int dt \left[\delta \Phi^{\ast}
T^{-1}(x)D_{x}^{(+)}\Phi\right]_{x}
\end{eqnarray}
Since $\delta\Phi^{\ast}=0$ at the boundary, requiring $\delta S=0$
implies that
\begin{equation}
-f(q)D_{x}^{(+)}T^{-1}(x)D_{x}^{(+)}\Phi=i \dot{\Phi}-V(x) \Phi ,
\end{equation}
which is equivalent to Equation (\ref{H}) provided that $f(q)=q^{1/2}$.  In
lattice notation, the lagrangian $L(t)$ can be alternatively written as
\begin{eqnarray}
L(t) & = & \sum_{n=0}^{\infty}L_{n}
    = \sum_{n=0}^{\infty}\Delta x_{n}
\left[-q^{1/2} \frac{\Delta\Phi^{\ast}(x_{n})}{\Delta x_{n}}
\frac{\Delta\Phi (x_{n})}{\Delta x_{n}}\right]\nonumber\\
 &  & + \sum_{n=0}^{\infty} \Delta x_{n}\left[i\Phi^{\ast}(x_{n})
\dot{\Phi}(x_{n})-V(x_{n})\Phi^{\ast}(x_{n})\Phi(x_{n}) \right] \label{L}
\end{eqnarray}
where we denoted $x_{n}\equiv q^{n}x$.
The momentum density conjugate to $\Phi$ is given
by
\begin{equation}
\pi(x_{m})=\frac{\partial{\cal L}_{m}}{\partial\dot{\Phi}(x_{m})}
=i \Phi^{\ast}(x_{m})
\end{equation}
where ${\cal L}_{m}\equiv \frac{L_{m}}{\Delta x_{m}}$. Similarly,
the hamiltonian becomes
\begin{equation}
H=\sum_{n=0}^{\infty}\Delta x_{n}\left[-iq^{1/2}\frac{\Delta\pi(x_{n})}
{\Delta x_{n}} \frac{\Delta\Phi(x_{n})}{\Delta x_{n}}-iV(x_{n})
\pi(x_{n})\Phi(x_{n})\right]
\end{equation}
Let us specialize to the free case, and consider $\Phi(x,t)$ as
a quantum field  restricted
to the region $0\leq x\leq d$.  A solution to Equation (\ref{H}) with
these boundary conditions can be written as a expansion
\begin{equation}
\Phi(x,t)=\sum_{\kappa} \frac{a(\kappa)}{\sqrt{M_{\kappa}}}
sin\left(\sqrt{q};\kappa x \right) e^{-i\epsilon_{\kappa} t} . \label{Phi}
\end{equation}
Since the differential operator $H_{0}$ is hermitian the conjugate field
$\overline{\Phi}(x,t)$ satisfies the same differential equation, and it
can be similarly written as
\begin{equation}
\overline{\Phi}(x,t)=\sum_{\kappa}
\frac{\overline{a}(\kappa)}{\sqrt{M_{\kappa}}}
sin\left(\sqrt{q};\kappa x \right) e^{i\epsilon_{\kappa} t}, \label{Phitilde}
\end{equation}
The energy eigenvalues are $\epsilon_{\kappa}=k\overline {k}\equiv \kappa^{2}$,
where the allowed $\kappa$ values satisfy $sin\left(\sqrt{q};\kappa d
\right)=~0$.
Since the operator $p$ and the hamiltonian $H_{0}$
do not commute they do not share the same
eigenfunctions. The variables $k$ and $\overline {k}$ denote the eigenvalues of
$p$ and $\overline {p}$ with eigenfunctions $e_{q}^{ikx}$ and
and $e_{q^{-1}}^{iq\overline{k}x}$ respectively. These functions
are defined in terms of the
Eulerian series
\begin{equation}
e_{q}^{x}=\sum_{n=0}^{\infty}\frac{x^{n}}{\{n\}!} \hspace{.1in} , \hspace{.1in}
 \{n\}=\frac{1-q^{n}}{1-q}
\end{equation}
\begin{equation}
sin\left(\sqrt{q}; x \right)=\sum_{n=0}^{\infty} (-1)^{n}
\frac{ x^{2n+1}}{[2n+1]!}
\end{equation}
The function $sin\left(\sqrt{q};\kappa x \right)$ satisfies the
orthogonality relation
\begin{equation}
\frac{1}{\sqrt{M_{\kappa}M_{\kappa'}}}\int_{0}^{qd} d_{q}x \,
sin\left(\sqrt{q};\kappa x \right)
sin\left(\sqrt{q};\kappa' x \right)= \delta_{\kappa,\kappa'} \label{or}
\end{equation}
with  $M_{\kappa}=\int_{0}^{qd} d_{q}x \left[
sin\left(\sqrt{q};\kappa x \right)\right]^{2}$, such that the
annihilation and creation operators can be written in terms of
the field $\Phi(x,t)$ as follows
\begin{eqnarray}
a(\kappa) & = & \int_{0}^{qd} \frac{d_{q}x}{\sqrt{M_{\kappa}}} \Phi(x,t)
sin\left(\sqrt{q};\kappa x \right) e^{i\epsilon_{\kappa}t} \nonumber \\
 & = & \sum_{n=0}^{\infty} \frac{(1-q)d_{n+1}}{\sqrt{M_{\kappa}}}
\Phi(d_{n+1},t) sin\left(\sqrt{q};d_{n+1}\kappa \right)
 e^{i\epsilon_{\kappa}t} \label{a}
\end{eqnarray}
and
\begin{eqnarray}
\overline{a}(\kappa) & = & \int_{0}^{qd} \frac{d_{q}x}{\sqrt{M_{\kappa}}}
\overline{\Phi}(x,t)
sin\left(\sqrt{q};\kappa x \right) e^{-i\epsilon_{\kappa}t} \nonumber\\
 & = & \sum_{n=0}^{\infty} \frac{(1-q)d_{n+1}}{\sqrt{M_{\kappa}}}
\overline{\Phi}(d_{n+1},t) sin\left(\sqrt{q};d_{n+1}\kappa \right)
 e^{-i\epsilon_{\kappa}t}. \label{abar}
\end{eqnarray}
where $d_{n}\equiv q^{n}d$.
Now, let the fields $\Phi$ and $\overline{\Phi}$ satisfy on the lattice
the deformed commutation relations
\begin{equation}
\Phi(x_{n},t) \overline{\Phi}(x_{m},t)-q^{2} \overline{\Phi}(x_{m},t)
\Phi(x_{n},t)= \frac{\delta_{n,m}}{\Delta x_{n}}  \label{qc1}
\end{equation}
which in the $q\rightarrow 1$ limit becomes $\left[\Phi(x,t),
\overline{\Phi}(x,t)\right]\rightarrow \infty$,
 and the commutation relation
\begin{equation}
\left[\Phi(x_{n},t),\Phi(x_{m},t)\right]=0.
\end{equation}
 After integration by parts
the hamiltonian operator $H_{0}$ becomes
\begin{equation}
H_{0}=-q^{1/2} \sum_{n=0}^{\infty} \Delta x_{n} \overline{\Phi}(x_{n})
\frac{\Delta}{\Delta x_{n}} T^{-1}\frac{\Delta\Phi(x_{n})}{\Delta x_{n}},
\end{equation}
such that with use of Equation (\ref{qc1}) we find that the time
evolution of the field $\Phi$ can be also written as
\begin{equation}
i\dot{\Phi}(x,t)=\Phi(x,t)H_{0}-q^{2}H_{0}\Phi(x,t),\label{qc2}
\end{equation}
where that for $\overline{\Phi}$ is given by the hermitian adjoint
equation.  The asymmetry introduced by the parameter $q$
does not allow  Equation (\ref{qc2}) to apply  for an arbitrary
function of the fields.  For example, Equation (\ref{qc2})
and its hermitian adjoint imply that the time evolution of the function
$F=\overline{\Phi}\Phi$ is written in terms
of a commutator
\begin{equation}
i\dot{F}=\left[F,H_{0}\right] ,
\end{equation}
and thus $\dot{H}_{0}=0$, as expected.
{}From Equation (\ref{qc1}) and the orthogonality relation in Equation
(\ref{or})
we find that the operators $a(\kappa')$ and $\overline{a}(\kappa)$
satisfy
\begin{equation}
\left[a(\kappa'),\overline{a}(\kappa)\right]_{q^{2}}\equiv
a(\kappa')\overline{a}(\kappa)-q^{2}\overline{a}(\kappa)a(\kappa')=
\delta_{\kappa,\kappa'}
\end{equation}
 In terms of  $a(\kappa')$ and $\overline{a}(\kappa)$ operators
the hamiltonian reads
\begin{equation}
H_{0}= \sum_{\kappa} \kappa^{2}\overline{a}(\kappa) a(\kappa).
\end{equation}
The vacuum expectation value between two fields is given by
\begin{eqnarray}
G(x,x',t,t') & \equiv & <\Phi(x,t) \overline{\Phi}(x',t')> \nonumber \\
& = &  \sum_{\kappa}\frac{1}{M_{\kappa}} sin\left(\sqrt{q};\kappa x \right)
sin\left(\sqrt{q};\kappa x' \right) e^{i\epsilon_{\kappa}(t'-t)}. \label{vev}
\end{eqnarray}
{}From Equations (\ref{or}) and (\ref{vev}) we obtain that
$G(x,x',t,t')$ relate the field
at two  lattice points $d_{n}$ and $d_{m}$ according to
\begin{equation}
\Phi(d_{m},t)=\sum_{n=0}^{\infty} (1-q)d_{n} G(d_{m},d_{n},t,t')
\Phi(d_{n},t'),
\end{equation}
and therefore
\begin{equation}
G(d_{m},d_{n},t,t)=\frac{\delta_{n,m}}{\Delta d_{n}},
\end{equation}
which itself represents a closure relation.
As a simple example, we consider the case $q\approx 0$.  A good approximation
to the zeros of the basic sine function is given by \cite{Exton}
\begin{equation}
\alpha_{m}\approx \frac{q^{-m+1/4}}{1-q}
\end{equation}
with $m=1,2,3,...$, and therefore Equation (\ref{vev}) can be written as a
expansion over the positive integers as follows

\begin{equation}
 <\Phi(x,t) \overline{\Phi}(x',t')>_{q\approx 0} =
\frac{q}{M_{1}} \sum_{n=1}^{\infty}
q^{-n} sin\left(\sqrt{q};\kappa_{n} x \right)
sin\left(\sqrt{q};\kappa_{n} x' \right) e^{i\kappa_{n}^{2}(t'-t)} \label{vev0}
\end{equation}
where we defined $\kappa_{n}\equiv\frac{q^{-n+1/4}}{(1-q)d}$, and the
coefficient
\begin{equation}
M_{\kappa=\kappa_{n}}\equiv M_{n}  =  \int_{0}^{qd} d_{q}x \,
sin^{2}\left(\sqrt{q};\kappa_{n} x \right)
 =  q^{n-1} M_{1}
\end{equation}
Now, once we  take two lattice points $x=d_{l}$ and $x'=d_{m}$, we see
that Equation (\ref{vev0}) reduces to a finite sum. For example, if
$l \leq m$ we have
\begin{equation}
 <\Phi(d_{l},t) \overline{\Phi}(d_{m},t')>_{q\approx 0}  =
 \frac{q}{M_{1}} \sum_{n=1}^{l}
q^{-n} sin\left(\sqrt{q};\kappa_{n}d_{l} \right)
 sin\left(\sqrt{q};\kappa_{n}d_{m} \right) e^{i\kappa_{n}^{2}(t'-t)} .
\end{equation}

\section{$SU_{q}(2)$ covariant field algebras} \label{fa}
In this section, based on the formalism given in Section \ref{QFD},
we first exhibit explicitly the covariant algebras satisfied by two $q$-bosonic
and $q$-fermionic fields.
Let us introduce two Schr\"{o}dinger  fields $\Phi_{1}$ and $\Phi_{2}$.
The free full  lagrangian is, of course, the sum of the two
corresponding lagrangians,as given in Equation (~\ref{L}) with $V=0$.
It is simple to check that this Lagrangian is invariant under
$SU_{q}(2)$ transformations
\begin{equation}
\Phi'_{i}=T_{ij}\Phi_{j}   \label{t}
\end{equation}
where the $SU_{q}(2)$ matrix is given by
\begin{equation}
T=\left(\begin{array}{cc} a & b \\ c & d\end{array}\right)
\end{equation}
and the adjoint matrix $\overline{T}$
\begin{equation}
\overline{T}=\left(\begin{array}{cc} d & -qb \\ -q^{-1}c & a\end{array}\right)
\end{equation}
with coefficients satisfying the well known relations
\begin{eqnarray}
ab=q^{-1}ba  & , & ac=q^{-1}ca \nonumber \\
bc=cb & , & dc=qcd  \nonumber \\
db=qbd & , &  da-ad=(q-q^{-1})bc  \nonumber \\
& & det_{q}T\equiv ad-q^{-1}bc=1  .
\end{eqnarray}
Now, a set of $q$-bosonic oscillators transforming
linearly as in Equation (\ref{t}) form a $SU_{q}(2)$ covariant algebra
according to the following relations
\begin{equation}
\left[a_{1}(\kappa'),\overline{a}_{1}(\kappa)\right]_{q^{2}}=
\delta_{\kappa',\kappa}+(q^{2}-1) \overline{a}_{2}(\kappa) a_{2}(\kappa')
\end{equation}
\begin{equation}
\left[a_{2}(\kappa'),\overline{a}_{2}(\kappa)\right]_{q^{2}}=
\delta_{\kappa,\kappa'}
\end{equation}
\begin{equation}
a_{2}(\kappa)a_{1}(\kappa')=q a_{1}(\kappa')a_{2}(\kappa)
\end{equation}
\begin{equation}
a_{2}(\kappa)\overline{a}_{1}(\kappa')=q \overline{a}_{1}(\kappa')a_{2}(\kappa)
,
\end{equation}
and according to Equations (\ref{or}), (\ref{a}) and (\ref{abar}) we
obtain that the
corresponding  $SU_{q}(2)$ covariant field algebra is given by
\begin{equation}
\left[\Phi_{1}(x_{n},t),\overline{\Phi}_{1}(x_{m},t)\right]_{q^{2}}=
\frac{\delta_{m,n}}{\Delta x_{n}}+(q^{2}-1)\overline{\Phi}_{2}(x_{m},t)
\Phi_{2}(x_{n},t)
\end{equation}
\begin{equation}
%% FOLLOWING LINE CANNOT BE BROKEN BEFORE 80 CHAR
\left[\Phi_{2}(x_{n},t),\overline{\Phi}_{2}(x_{m},t)\right]_{q^{2}}=\frac{\delta_{m,n}}{\Delta x_{n}}
\end{equation}
\begin{equation}
\Phi_{2}(x_{n},t)\Phi_{1}(x_{m},t)=q\Phi_{1}(x_{m},t)\Phi_{2}(x_{n},t)
\end{equation}
\begin{equation}
\Phi_{2}(x_{n},t)\overline{\Phi}_{1}(x_{m},t)=
q\overline{\Phi}_{1}(x_{m},t)\Phi_{2}(x_{n},t)
\end{equation}
Similarly, starting from a $SU_{q}(2)$ covariant $q$-fermionic algebra
\begin{eqnarray}
b_{1}(\kappa)b_{2}(\kappa')=-q b_{2}(\kappa')b_{1}(\kappa) & , &
\overline{b}_{1}(\kappa)b_{2}(\kappa')=-q b_{2}(\kappa')
\overline{b}_{1}(\kappa)
\end{eqnarray}
\begin{equation}
\{b_{1}(\kappa),\overline{b}_{1}(\kappa')\}=
\delta_{\kappa,\kappa'}-(1-q^{-2})\overline{b}_{2}(\kappa')
b_{2}(\kappa)
\end{equation}
\begin{equation}
\{b_{2}(\kappa),\overline{b}_{2}(\kappa')\}=
\delta_{\kappa,\kappa'} ,
\end{equation}
the set of $q$-fermionic fields $\Psi_{1}$ and $\Psi_{2}$
satisfying Equation (\ref{H}), with $V=0$, with expansion
\begin{equation}
\Psi_{j}(x,t)=\sum_{\kappa} \frac{b_{j}(\kappa)}{\sqrt{M_{\kappa}}}
sin\left(\sqrt{q};\kappa x \right) e^{-i\epsilon_{\kappa} t} \; , \; j=1,2
\end{equation}
satisfy the following $SU_{q}(2)$ covariant field algebra
\begin{equation}
\Psi_{1}(x_{n},t)\Psi_{2}(x_{m},t)=-q \Psi_{2}(x_{m},t)\Psi_{1}(x_{n},t)
\end{equation}
\begin{equation}
\overline{\Psi}_{1}(x_{n},t)\Psi_{2}(x_{m},t)=-q \Psi_{2}(x_{m},t)
\overline{\Psi}_{1}(x_{n},t)
\end{equation}
\begin{equation}
\{\Psi_{1}(x_{n},t),\overline{\Psi}_{1}(x_{m},t)\}=
\frac{\delta_{m,n}}{\Delta x_{n}} - (1-q^{-2})
\overline{\Psi}_{2}(x_{m},t)\Psi_{2}(x_{n},t)
\end{equation}
\begin{equation}
\{\Psi_{2}(x_{n},t),\overline{\Psi}_{2}(x_{m},t)\}=
\frac{\delta_{m,n}}{\Delta x_{n}}
\end{equation}
\begin{equation}
\{\Psi_{1}(x_{m},t),\Psi_{1}(x_{n},t)\}=0=
\{\Psi_{2}(x_{m},t),\Psi_{2}(x_{n},t)\}
\end{equation}
The operators $\overline{a}_{j}$ , $a_{j}$ and $\overline{b}_{j}$,
for $\kappa=\kappa'$ and $j=1,2$ can be
put in one to one correspondence with the quantum plane coordinates,
differentials and derivatives in Ref. \cite{WessZumino}.
The generalization of these relations
to $n$ $q$-bosonic and $q$-fermionic fields follows.  In
compact form, we write
\begin{equation}
\Omega_{j}(x_{n})\overline{\Omega}_{i}(x_{m})=
\delta_{i,j}\frac{\delta_{m,n}}{\Delta x_{n}}\pm q^{\pm 1}
\hat{{\cal R}}_{ikjl}(q)
\overline{\Omega}_{l}(x_{m})\Omega_{k}(x_{n})
\end{equation}
\begin{equation}
\Omega_{i}(x_{n})\Omega_{j}(x_{m})=\pm q^{\mp 1}  \hat{{\cal R}}_{lkji}(q)
\Omega_{k}(x_{m})\Omega_{l}(x_{n})
\end{equation}
\begin{equation}
\overline{\Omega}_{i}(x_{n})\overline{\Omega}_{j}(x_{m})=
\pm q^{\mp 1}  \hat{{\cal R}}_{ijkl}(q)
\overline{\Omega}_{k}(x_{m})\overline{\Omega}_{l}(x_{n})
\end{equation}
where $\Omega$ denotes either $\Phi$ or $\Psi$, and the upper
(lower) sign applies to $\Phi$ $(\Psi)$ field. The matrix
$ \hat{{\cal R}}_{ijkl}(q)$ is given by
\begin{equation}
 \hat{{\cal R}}_{ijkl}(q)=R_{jikl}(q^{-1}),
\end{equation}
where $R_{ijkl}(q)$ is the $R$-matrix of $\hat{A}_{n-1}$ \cite{Takhtajan}
.  There is
an additional covariant algebra involving $q$-bosonic and
$q$-fermionic fields.  In terms of the matrix $\hat{{\cal R}}$ it reads
\begin{equation}
\overline{\Phi}_{i}(x_{n})\overline{\Psi}_{j}(x_{m})=
q  \hat{{\cal R}}_{ijkl}(q) \overline{\Psi}_{k}(x_{m})
\overline{\Phi}_{l}(x_{n})
\end{equation}
\begin{equation}
\overline{\Psi}_{i}(x_{n})\Phi_{j}(x_{m})=q
 \hat{{\cal R}}_{ljki}(q) \Phi_{l}(x_{m})
\overline{\Psi}_{k}(x_{n})
\end{equation}
\begin{equation}
\Psi_{i}(x_{n})\Phi_{j}(x_{m})=q  \hat{{\cal R}}_{lkji}(q)
\Phi_{k}(x_{m})\Psi_{l}(x_{n})
\end{equation}
\section{Discussion}
In this letter, starting from the equivalence of a quantum deformation of
quantum mechanics
and quantum mechanics on a nonlinear lattice,
we developed the basic formalism of a $q$-bosonic ($q$-fermionic)
 Schr\"{o}dinger quantum field theory
with the fields satisfying quantum commutation (anticommutation)
 relations on the quantum line .
We have seen that the free field case admits a field expansion in terms of
orthogonal functions, in a similar way as in undeformed  Schr\"{o}dinger
field theory.
Then,  based on this formalism, we considered $SU_{q}(2)$ as
an internal quantum symmetry and obtained the  covariant  algebra
satisfied by the
fields, and generalized this result to an arbitrary number of fields.
The formulation we have proposed in this work  gives an additional
geometrical insight into the meaning of quantum deformations and it
could give a new angle to approach
quantum field theory  in non-commutative geometry.
There are several proposals of either quantum group field theory or
$q$-deformed field theory,  most
of them on classical space-time, which can be found in the literature
\cite{Arefeva}.
It would be interesting to generalize some of the issues addressed in this
paper to more general situations involving
a nonzero potential and  higher number of quantum dimensions.

\end{document}